\documentclass{aa}
\usepackage{graphicx}
\usepackage{txfonts}
\begin{document}
   \title{A deep, wide-field search for substellar members in NGC\,2264}
    
     \author{T.R. Kendall
          \inst{1}
          \and
          J. Bouvier\inst{1}
          \and
          E. Moraux\inst{2}
          \and
          D.J. James\inst{1,3}
          \and 
          F. M\'{e}nard\inst{1}
          }

   \offprints{T.R. Kendall}

   \institute{Laboratoire d'Astrophysique, Observatoire de Grenoble, Universit\'{e} 
              Joseph Fourier, F-38041 Grenoble Cedex 09, France\\
              \email{tkendall@obs.ujf-grenoble.fr, jbouvier@obs.ujf-grenoble.fr}
         \and
             Institute of Astronomy, University of Cambridge, Madingley Road, Cambridge CB3 0HA, United Kingdom\\
             \email{moraux@ast.cam.ac.uk}
         \and
             Department of Physics and Astronomy, Vanderbilt University, 1807 Station B, Nashville, TN 37235, USA
   \thanks{Based on observations made with the Canada-France-Hawaii telescope and data from the 2MASS
           project (University of Massachusetts and IPAC/Caltech, USA.)}
             }

   \date{Received xxx; accepted xxx}

\abstract{We report the first results of our ongoing campaign to discover the first brown dwarfs (BD) in NGC\,2264, 
a young (3\,Myr), 
populous star forming region for which
our optical studies have revealed a very high density of potential candidates - 236 in $<$ 1\,deg$^2$ - 
from the substellar limit down to at least $\sim$\,20\,M$_{\rm Jup}$ for zero reddening. Candidate BD
were first selected using wide field ($I,z$) band imaging with CFH12K, by reference to 
current theoretical isochrones. Subsequently, 79
(33\%) of the $I,z$ sample were found to have
near-infrared 2MASS photometry ($JHK_s$\,$\pm$\,0.3\,mag. or better), yielding dereddened magnitudes and allowing further investigation by
comparison with the location of NextGen and DUSTY isochrones in colour-colour and colour-magnitude
diagrams involving various combinations of $I$,$J$,$H$ and $K_s$.  We discuss
the status and potential substellarity of a number of relatively unreddened (A$_{\rm v}$\,$\la$\,5) likely low-mass members in our sample, 
but in spite of the depth of our observations in $I,z$, we 
are as yet unable to unambiguously identify substellar candidates using only 2MASS data. Nevertheless, there are excellent 
arguments for considering two faint (observed $I$\,$\sim$\,18.4 and 21.2) objects as cluster candidates with masses respectively at or rather below the
hydrogen burning limit. More current candidates 
could be proven to be cluster members with masses around 0.1\,M$_{\odot}$ {\it via} gravity-sensitive spectroscopy, and deeper near-infrared imaging
will surely reveal a hitherto unknown population of young brown dwarfs
in this region, accessible to the next generation of deep near-infrared surveys. 
               }

   \maketitle
%

\section{Introduction}

Within the last few years the observational study of substellar objects (M\,$<$0.072\,M$_{\odot}$) has undergone
spectacular and rapid development, revealing new perspectives on the formation of such objects within molecular clouds.
Large numbers of brown dwarfs (BD) have now been found in star-forming regions (SFRs), young clusters and the 
field (\cite{bej01}; \cite{mor03}; \cite{cha02}; \cite{cha03}; \cite{cru03}) and physical models of the atmospheres of BD 
constructed (\cite{bar03}; \cite{cha00}), opening up the possibility of confrontation between observations and 
theoretical predictions of the physical properties of BD. 
However, the core issues of the form of the substellar initial mass function (IMF) 
and its dependence on
environment remain to be addressed. The discovery of BD in widely differing environments does suggest their formation is
directly linked to the star formation process, and early estimates of the substellar IMF in the solar neighbourhood 
have indicated that BD are nearly as numerous as stars. Within this broad framework, two competing scenarios have been
put forward for BD formation; the first simply that they form as stars do, i.e. by the gravitational collapse of low mass
molecular cloud cores (\cite{pad02}); the second postulates the dynamical ejection of the lowest mass protostars, leading to
BD formation since the ejected fragments are unable to further accrete (\cite{rei01}). To distinguish between these
possibilities and to obtain unbiased estimates of the substellar IMF, observations of statistically complete, homogeneous
populations of BD are needed in a wide variety of environments with different ages.

 Young star-forming regions
are of particular interest since the youth of such objects  ensures that their population has not yet
suffered from important dynamical (and stellar) evolution, other than dynamical effects 
potentially associated with their formation. This is a point worth underlining, since observations of
young SFRs have the potential to distinguish these two BD formation mechanisms
by their resultant spatial distributions. If BD always form exactly as stars do, one would expect them to
trace the regions of the highest density of low-mass stars. However, in any dynamical ejection scenario,
a deficit of BD may be expected in the central regions of young star-forming clusters. Their initial
velocities of a few km\,s$^{-1}$ (\cite{ste03}) would especially imply a preferential loss of BD from 
loosely bound regions such as Taurus, where a velocity of 1\,km\,s$^{-1}$ would correspond to a drift
of $\sim$\,0.3$^{\circ}$/\,Myr at 140\,pc (or in absolute terms $\sim$\,0.7\,pc\,/\,Myr). In denser
environments, BD would remain bound, but \cite{mor04} found that BD would still be expected to have a
more extended distribution than the stars, in clusters whose age is comparable to the crossing time.
Such a difference in spatial distribution should be observable in all nearby such SFRs; 
we note also that other models predict rather similar dispersion velocities for BD and more massive members (\cite{bat03}). Wide field
observations of large BD samples in these regions will soon settle this issue. 

Young BD (with ages 1--3\,Myr) are now being uncovered (for a relatively recent compilation see \cite{bas00}), but 
the census remains very incomplete to date. Comparison between the BD populations of young SFRs with a range of environmental conditions
permits investigation of the sensitivity of the low-mass
end of the IMF to local conditions, and other  
more recent studies are beginning to address 
this question. 
To cite two specific cases, 
30 substellar candidates have been identified in Taurus in a 3.6\,deg$^2$ region using CFH12K
data which reached to $I$\,=\,23.5. Spectroscopic follow-up led to the identification of 
four BD in Taurus with spectral type later than M7 (\cite{mar01}). In Taurus, $\sim$10 BD are
now known (\cite{bri02}), and appear to be spatially correlated with regions of highest  stellar
density. In Upper Scorpius, initial studies (\cite{mar04}) have found 18 candidate BD of which 5
show signs of ongoing accretion. 

In this paper, we present initial results from a deep CFH12K survey of the young (3\,Myr), rich
star-forming region NGC\,2264.
The study introduced here is only a small part (0.6\,deg$^2$) of a much broader survey, 
which has mapped
significant areas (70\,deg$^2$) of a variety of environments (ages
1--600\,Myr) sufficiently deeply to probe BD; in all of the targeted regions, the limiting mass is in the
range 10--40\,M$_{\rm Jup}$. NGC\,2264, because of its youth and relative proximity 
(3\,Myr, 760\,pc; \cite{lam04}; \cite{reb02}; \cite{sun97}; \cite{par00}), is probed to the lower
end of this mass range with the current observations. We will discuss our initial findings on NGC\,2264 in Sect.\,4;
here we will note that the much larger coeval BD populations currently being uncovered by CFH12K, when fully
analysed, promise a more robust statistical treatment of both the kinematics of BD populations and variations
in the substellar IMF from region to region.

\begin{figure}
\centering
\includegraphics[width=9cm]{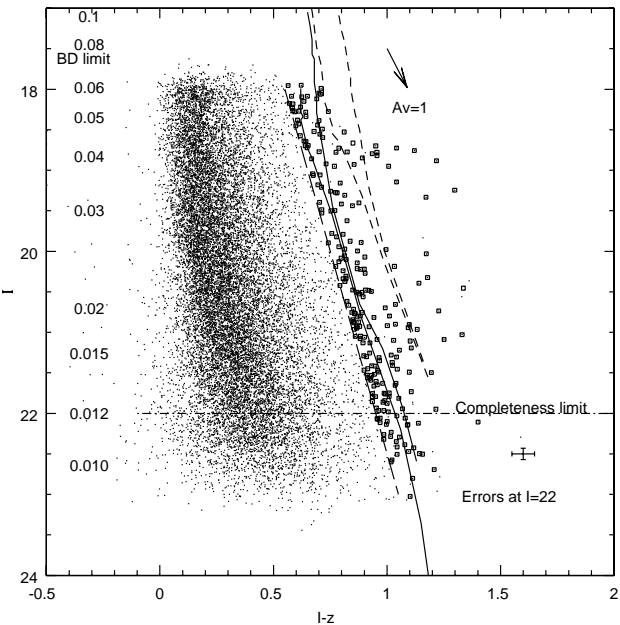}
\caption{$I$,\,$I-z$
colour-magnitude diagram built using long (3\,$\times$\,360\,sec) exposures of NGC\,2264.
The solid isochrones are DUSTY models for ages of 2 and 5\,Myr; the dashed isochrones
are the NextGen models for the same ages. All isochrones are for distance modulus 9.4; 760\,pc. 236
BD candidates (squares)
were selected to be redward of the sloping straight line (long dash). Small dots
 redward of this
line were rejected from the candidacy on visual inspection. Typical error bars 
for $I$\,=\,22 of $\sim$\,0.1\,mag are shown. 
The mass scale 
(in M$_{\odot}$) is for the 2\,Myr models; it  and the estimated completeness limit are indicated.}
\label{f1}
\end{figure} 

\section{Observations of candidate NGC\,2264 members: Optical colour selection}

Observations of NGC\,2264 were carried out using the Canada-France-Hawaii Telescope (CFHT) and the 12K camera, with a mosaic
of 12 4128\,$\times$\,2080 15\,$\mu$m pixel CCDs as detector yielding a field of view of 43$^{\prime}$\,$\times$\,28$^{\prime}$ with
0.206$\arcsec$\,pix$^{-1}$.
2 fields were observed in the $I$-band on 2 February 1999, and the same fields covered in the $z$-band on 19 December 2000; in both cases
each observation is a composite of 3\,$\times$\,360\,second exposures. Weather conditions were good and the seeing 1.2$\arcsec$ or better.
The pre-reduction and analysis of these data have been performed using CFHT Elixir\,\footnote{{\tt www.cfht.hawaii.edu}}
pipelines and new, innovative point-spread function (PSF) fitting techniques (SExtractor, PSFex) developed at the
Institut d'Astrophysique de Paris by E. Bertin (\cite{ber96})\,\footnote{see {\tt www.terapix.iap.fr} for more recent developments}. 
After pipeline image flat-fielding and co-addition, $z$-band images (taken at a later epoch) were
mapped on to the $I$-band images using standard techniques ({\sc geomap, geotrans}) within the 
IRAF\,\footnote{IRAF is distributed by the National Optical Astronomy Observatories, which is operated by
the Association of Universities for Research in Astronomy, Inc. (AURA) under cooperative agreement with the National Science
Foundation} environment. Photometric extraction  was initially performed on the $I$-band images, with
subsequent matching to the $z$-images with an accuracy of $\pm$\,5\,pixels. The assessment of photometric
zeropoints at $I$ was
provided using images of the Landolt standard field SA\,101 (\cite{lan92}), observed with the same instrumental setup
at the same epoch. Where there were sufficient standard stars in each individual CCD of  the array (around half the 12 CCDs)
CCD-to-CCD variations in the zeropoint could be estimated, compared with the Elixir pipeline values (no discrepancies
larger than 0.1--0.2 mag. were found) and taken into account in our analysis. For the other $I$-band observations, zeropoints were
taken directly from the night-by-night CFHT/ Elixir database; this was also the case for the later-epoch $z$-band
observations, which are calibrated assuming $I-z$\,=\,0 for A0 stars. 
We estimate our typical random photometric errors to be $\pm$\,0.1 mag. at $I$\,=\,23 and $\pm$\,0.05
mag. at $I$\,=\,22, allowing us to place typical error bars in Fig.\,\ref{f1} which pertain near the completeness limit.

We performed a number of sample refining and cleaning procedures as follows; firstly; photometric points found to have
undefined magnitudes following PSF-fitting were rejected; this accounts for many detections close to bad columns, in diffraction spikes, 
etc. Candidates flagged to be 
saturated and/or near field edges were then also removed. At this point, the statistics of the remaining detections, placed in one-magnitude
bins in $I$, suggested that the data should be complete to very nearly $I$\,=\,23, there being very similar numbers of objects in this
faintest bin and one magnitude brighter. Subsequently, we retained only objects with a fraction-of-light radius (``flux radius'') within
quite narrow ranges estimated by eye by looking at plots of this quantity {\it versus} magnitude. 
For example, in $I$ only objects with a value of between 2 and 3.5 pixels were retained.
Furthermore, we rejected objects with an ellipticity of greater than 0.15, in a very efficient attempt to
rid our sample of extended or non-stellar objects. (To illustrate this, only one of 236 visually inspected candidates failed
the ellipticity cut; it is clear that this cut is
an efficient way of refining the sample, avoiding individual inspection of spurious points). 
As a final check on our selection, 
all 236 such $I,z$ candidates 
(squares in Fig.\,\ref{f1}) were carefully inspected visually
on both $I$ and $z$ frames and all found to be stellar in nature, i.e. there are no artifacts
due to nebulosity, field edges, bad columns, bright stars etc.
After all these steps, we retained all 236 candidates but only plot field objects
satisfying $e$\,$<$\,0.15, and all other criteria, in Fig.\,\ref{f1}. (Hence, the one candidate failing
this criterion is the only point
in Fig.\,\ref{f1} to be plotted as an open square with no inner dot).  
We find that the number of objects with $I$\,=\,22--23
is roughly 50\% of the number in the $I$\,=\,21--22 range; hence a completeness limit in the {\it refined} sample of $I$\,$\sim$\,22
is reasonable. We note that we made no attempt to remove candidates on the basis of their magnitudes only; our cuts have
rejected no fainter candidates, but a much larger number of points which prove, for one reason or another, clearly spurious
on inspection.  
Finally, astrometric calibration was carried out for each individual candidate using the Starlink
{\sc astrom} package. With an initial selection of $\sim$\,10 stars within the same CCD as the candidate, and input positions taken from the
USNO\,A\,2.0 catalogue, we found RMS accuracies 
of the initial fits to better than 0.1$\arcsec$.  

Fig.\,\ref{f1}. shows optically selected candidates in NGC\,2264 between the substellar limit and $\sim$\,10M$_{\rm Jup}$, 
by comparison with the state-of-the-art DUSTY models, specially created by Baraffe et al. to take account of the 
CFHT $I,z$ filter responses (see \cite{bar03} and references therein for full details of these models). 
In order to account for uncertainties in the model loci, and remain conservative in our candidate selection, all objects
redward of the sloping dashed line in Fig.\,\ref{f1} were retained for further investigation.  
As can also be seen in Fig.\,\ref{f1}, our estimate of
completeness to $I$\,$\sim\,$22, or 12\,M$_{\rm Jup}$ for age 2\,Myr (unreddened), appears accurate.

\begin{figure}
\centering
\includegraphics[width=9cm]{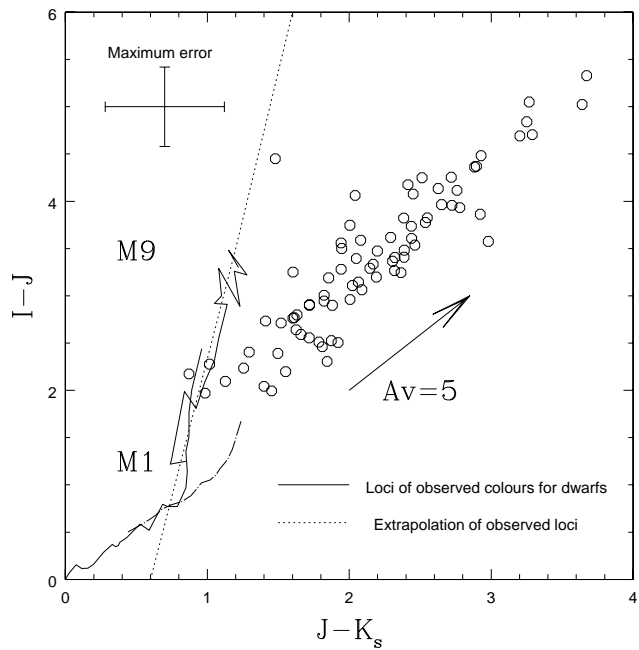}
\caption{All 79 $IzJHK_s$ candidates (open circles), plotted in the $I-J,J-K_s$ diagram. The errorbar assumes
$JHK_s$ accurate to $\pm$\,0.3\,mag and $I$ to $\pm$\,0.1\,mag. The reddening vector is shown.  
The solid lines are observed loci of dwarfs taken from a compilation of
sources (see text). The dash-dot line is a the locus of G0 -- M5 giants from \cite{bb88}. 
A$_{\rm v}$ has been computed by dereddening all points to the extrapolated (dotted) locus of dwarfs. 
After dereddening, all objects are potentially either cluster or field M-dwarfs.  }
\label{f2}
\end{figure}

\begin{figure}
\centering
\includegraphics[width=9cm]{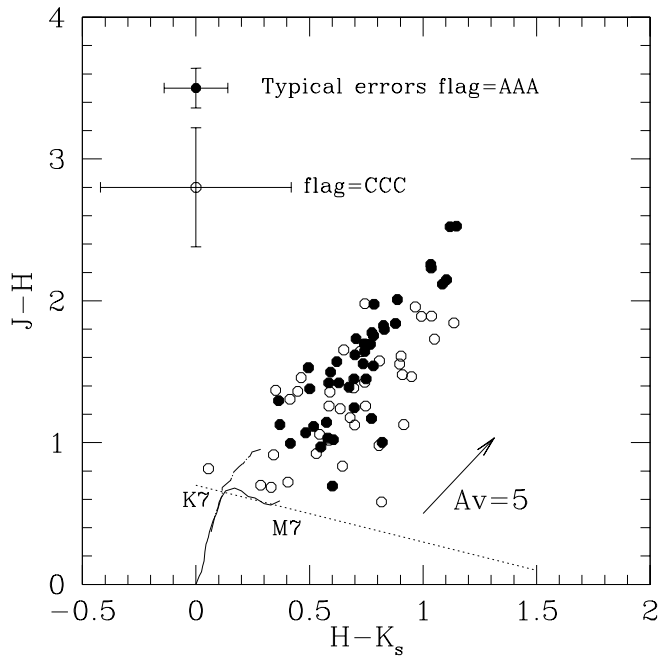}
\caption{Similar to Fig.\,\ref{f2}, but for $JHK_s$. Since the colours plotted here vary relatively less among the sample, the
errorbars appear larger. Therefore, a distinction has been made between the best 2MASS photometry (all magnitudes
flagged A, filled points) and those with any magnitude flagged B or C. For the former, $JHK_s$ are taken accurate to
$\pm$\,0.1\,mag., for the latter, $\pm$\,0.3\,mag., hence the errorbars plotted. Again, A$_{\rm v}$ has been calculated by dereddening
to the extrapolated (dotted) locus of observed late-type dwarf colours. The locus of giant colours is also plotted (dash-dot line). }
\label{f3}
\end{figure}

\begin{figure}
\centering
\includegraphics[width=9cm]{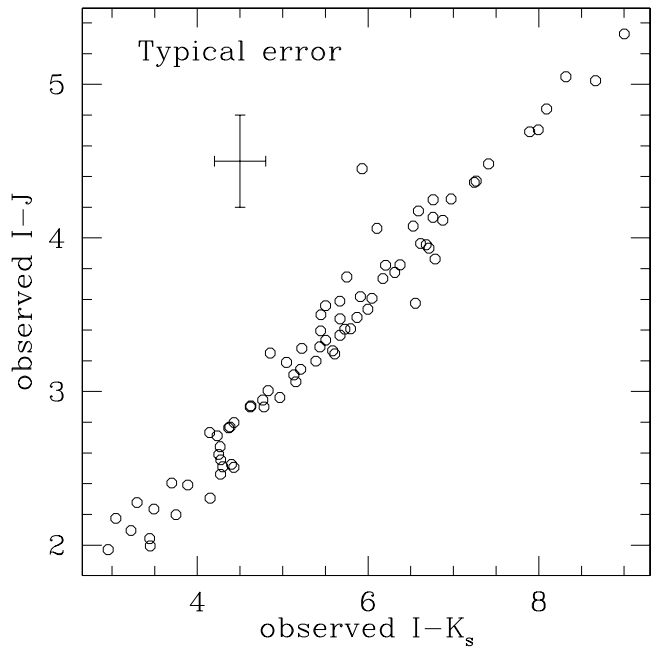}
\caption{To check for obvious $K_s$-band excesses which may affect our derivation of A$_{\rm v}$, we have plotted all 79 candidates
in the $I-J,I-K_s$ diagram. Clearly, no objects have significant excesses.  }
\label{f2a}
\end{figure}

\begin{figure}
\centering
\includegraphics[width=9cm]{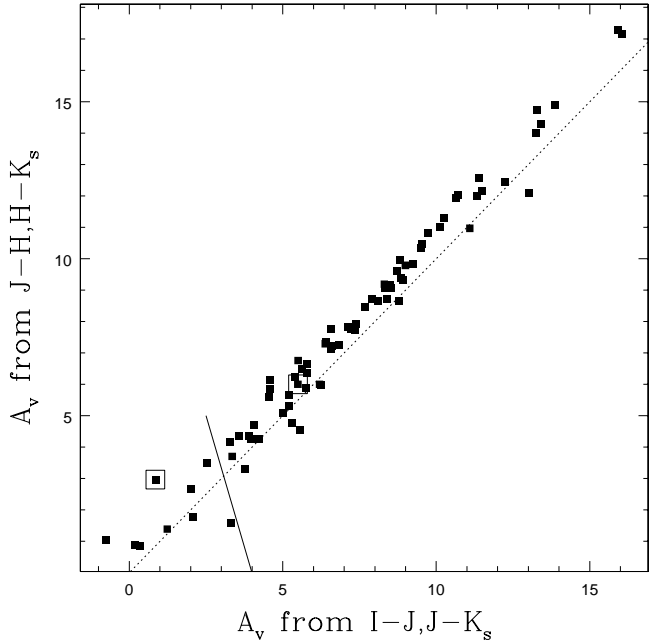}
\caption{Comparison of A$_{\rm v}$ derived by the two different methods indicated by the axis labels and detailed in the text. Both 
methods yield very similar values. The dotted line denotes equal A$_{\rm v}$ from each method. 
We have singled out eight objects with low reddening, lying to the left of the sloping line,
plus two more objects highlighted by open squares, for particular  discussion.  }
\label{f3a}
\end{figure}

\begin{figure}
\centering
\includegraphics[width=9cm]{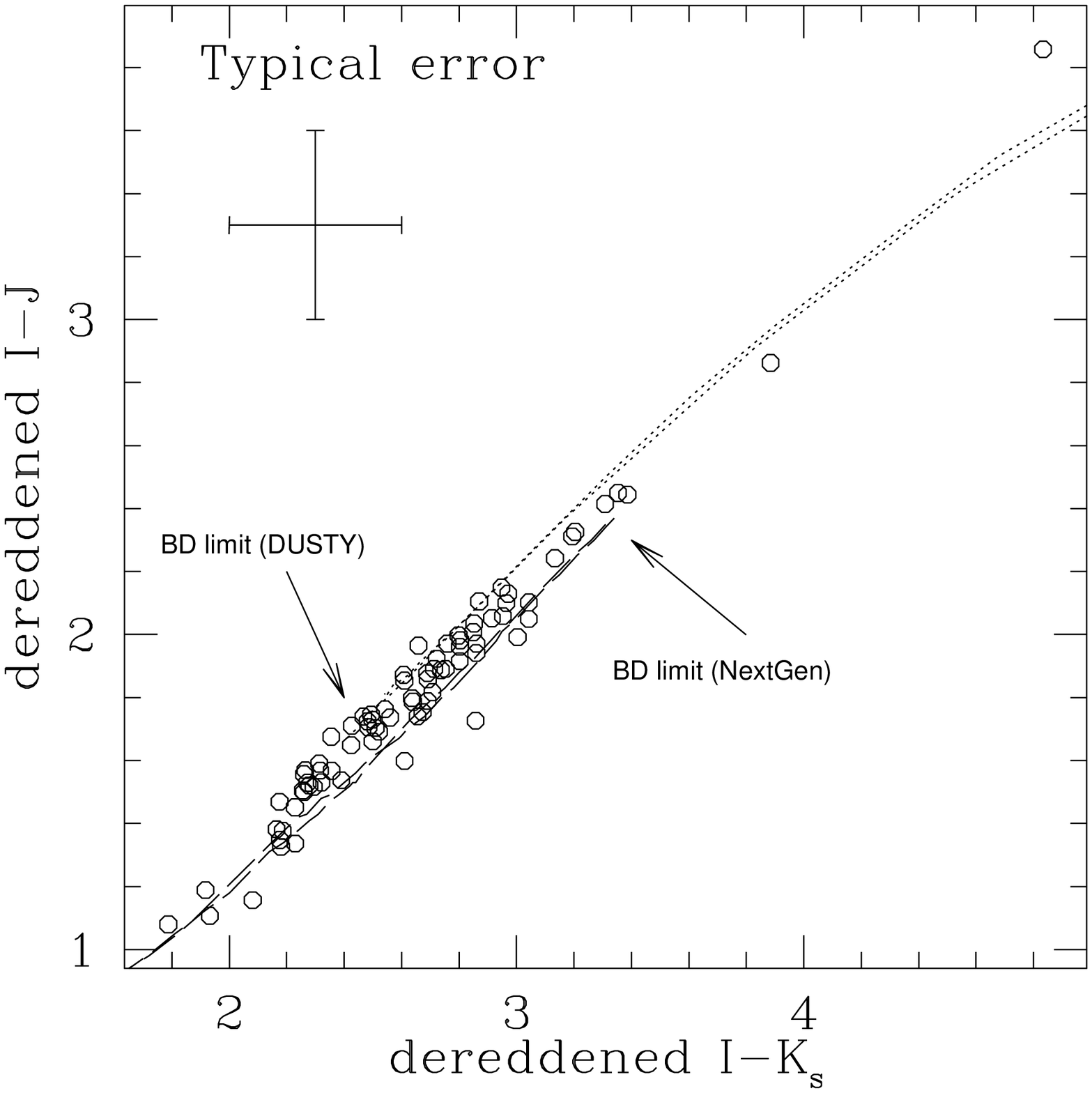}
\caption{The $I-J,I-K_s$ diagram, after dereddening. All objects lie close to the theoretical DUSTY and/or NextGen isochrones (dotted
and dashed lines, respectively). The substellar limit for suggested by each set of models is indicated; with reference to DUSTY models,
many objects are potentially substellar. The typical errors are the same as in Fig.\,\ref{f2a}.    }
\label{f2b}
\end{figure}

\section{Dereddening the sample}

In this section, we discuss our methods to estimate the extiction towards each member of our candidate sample. 
A brief examination of the reddening 
derived by previous authors is relevant at this point. At the galactic
coordinates for NGC\,2264 given by \cite{reb02}, ($l$\,=\,202.96, $b$\,=\,+2.22) the maps of \cite{sch98} yield A$_{\rm v}$\,=\,5.63, but it is
considered that a background dark molecular cloud exists (\cite{reb02}), limiting contamination by background giants. Indeed, these authors
find a typical extinction towards NGC\,2264 members A$_{\rm v}$\,$\sim$\,0.45 (from photometry and spectroscopy of late-type members), 
in good agreement with earlier values derived using small numbers of OB stars (\cite{sun97}; \cite{par00}). Noting that the
foreground reddening is effectively zero (\cite{sun97}), we therefore assume that A$_{\rm v}$ toward any given object in our sample
may vary from zero to high values.
Futhermore, with such large and variable extinction intrinsic
to NGC\,2264, it is clear that there may be relatively unreddened objects (perhaps on the nearer side of the cloud), and more
deeply embedded objects.

We find variable extinction values ranging up to A$_{\rm v}$\,$\sim$\,15, but generally in the range 0--10. Our methods rely on the
recovery of 79 objects in our sample of 236 (33\%) in the 2MASS database\,\footnote{{\tt www.ipac.caltech/edu/2MASS}}, where 2MASS
detections were required to be within 2\,\arcsec~ of our optical positions. More 2MASS
counterparts exist, but with either $J$- or $K_s$-magnitudes only given as upper limits - we do not consider these objects
further. The 79 objects will form the basis of the current study; all have photometric flags A, B or C, corresponding to errors
$\pm$\,0.1\,mag (A) to $\pm$\,0.3\,mag (C).

The basis for dereddening is given in Figs.~\ref{f2} and \ref{f3}. We have used observed loci for late-type dwarfs from compilations
given by \cite{leg92}, \cite{leg98}, \cite{bes91} and  \cite{kir00} to 
compare the location of our sample in the $I-J,J-K_s$ and  $J-H,H-K_s$ colour-colour diagrams.
In order to verify our methods we first checked our sample for obvious $K_s$-band excesses, which would strongly affect estimates
of the extinction which used the $K_s$-magnitude. We found none (Fig.\,\ref{f2a}).  

We plotted the location of
our sample in $I-J,J-K_s$, in which diagram the location of an unreddened M-dwarf locus can be easily described by a straight line 
(Fig.\,\ref{f2}).    
From the known slope of the reddening vector (\cite{rl85}) and the gradient of the line (dotted in Fig.\,\ref{f2}) the magnitude of
the vector A$_{\rm v}$ can be calculated {\it via} simple geometric arguments. 
Exactly similar arguments can be applied in the case
of Fig.\,\ref{f3}, where only near-infrared colours are utilised. In this case, the photometric errors are larger, with respect to
the range of $JHK_s$ colours observed. Hence, in Fig.\,\ref{f3}, we make a clear distinction between those objects (42 out of 79) which
have precise 2MASS photometry and those for which the errors are likely larger than $\pm$\,0.1\,mag. We find that the values of A$_{\rm v}$ 
found by these two methods are almost identical, aside from a few objects for which A$_{\rm v}$ is low and the
difference between the value of the extinction given by the two methods is of order  $\sim$\,1\,mag. (Fig.\,\ref{f3a}), which is not
a significant difference in the context of this work. Hence, in order to retain the important $I$-band magnitude for further analysis, 
we prefer to adopt
A$_{\rm v}$ given by the $J-H$ and $H-K_s$ colours, throughout the remainder of this paper.
We note here also, that in addition, we
performed similar experiments using a locus in the $I-K_s$,$J-H$ diagram, as adopted by \cite{luh03} for a study of the Taurus SFR, 
and found very similar values of 
A$_{\rm v}$ in excellent agreement with our preferred methods. Lastly, we are aware that dereddening to dwarf loci may not be 
correct for NGC\,2264 at age 3\,Myr, since young objects have surface gravities intermediate between dwarfs and giants (\cite{gl96}).
To illustrate this uncertainty, we have plotted the loci of giant stars of spectral type G0--M5 from the data of \cite{bb88} as
dash-dot lines in Figs.\,\ref{f2} and \ref{f3}. The plotted giant sequence ends at M5. With reference to Fig.\,\ref{f2}, one might 
assume that it would continue roughly parallel to the dwarf sequence; we do not have data to confirm this. Were it the case, however,
one can see from inspection that dereddening to a locus intermediate between dwarfs and giants might yield A$_{\rm v}$
systematically lower than our estimates, by about one magnitude. Hence our dereddened $I$-magnitudes would be fainter (by about 0.5\,mag.) 
and closer to the
observed values, and faint candidate members would therefore be more likely to be substellar with reference to theoretical models. One might 
also argue that, from Fig.\,\ref{f3a}, the values of A$_{\rm v}$ derived from the $J-H,H-K_s$ colour lie above the line of equal A$_{\rm v}$
with respect to the values from $I-J,J-K_s$, again by about 1 mag. Hence by choosing to use A$_{\rm v}$ derived from $JHK$ colours, we might 
have {\it overestimated} A$_{\rm v}$ by $\sim$\,1\,mag, thus A$_{\rm I}$ by $\sim$\,0.5\,mag. 
In that case, the true values of dereddened $I$ may be fainter by $\sim$\,0.5\,mag., and the masses correspondingly lower.
For these reasons, our derivation of dereddened magnitudes is conservative.

It is worth noting here that Fig.\ref{f2a} suggests
that we do not observe $K$-band excesses, which generally suggest the presence of an accretion disk. We have re-plotted this diagram in 
Fig.\,\ref{f2b}, 
for dereddened colours. 
The reader is encouraged to compare Fig.\,\ref{f2b} with the Fig.\,3 of \cite{cab04}, which shows 
the same plot for a number of $\sigma$\,Orionis young brown dwarfs, only one of which shows a clear $K_s$-band excess. For NGC\,2264, it is
not possible to disentangle the effect of the intrinsic reddening with the effects of accretion phenomena using 
near-infrared colours only; observations at longer wavelengths are required.

\begin{figure}
\centering
\includegraphics[width=9cm]{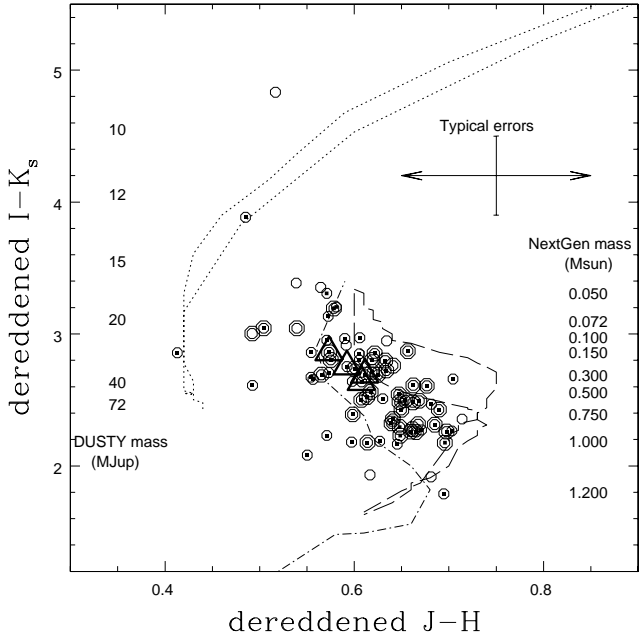}
\caption{The sample, plotted in the $I-K_s,J-H$ diagram, after dereddening. The dotted lines represent DUSTY isochrones for ages 
1 and 5\,Myr; the
dashed lines are NextGen models for 2 and 5\,Myr. The dash-dot line is a locus of observed dwarf colours. The errorbar in $J-H$ is large,
but mass scales for the mean of each pair of isochrones are derived from the much more diagnostic $I-K_s$ colour, and are shown for
DUSTY models (at left) and NextGen (at right). Note that the errorbars do not reflect uncertainties in A$_{\rm v}$ itself in this or subsequent figures 
(until
Fig.\,\ref{f10}, see Sect. 4.3). The whole sample is represented by open circles; points with the best 2MASS photometry 
(all magnitudes flagged A) are doubly overplotted
open symbols. The small filled squares are 65 {\it candidate} late-type cluster members (\cite{lam04}) recovered in our sample (see text); 
bold triangles are confirmed variable
PMS objects drawn from the same sample by these authors. }
\label{f4}
\end{figure}

\begin{figure}
\centering
\includegraphics[width=9cm]{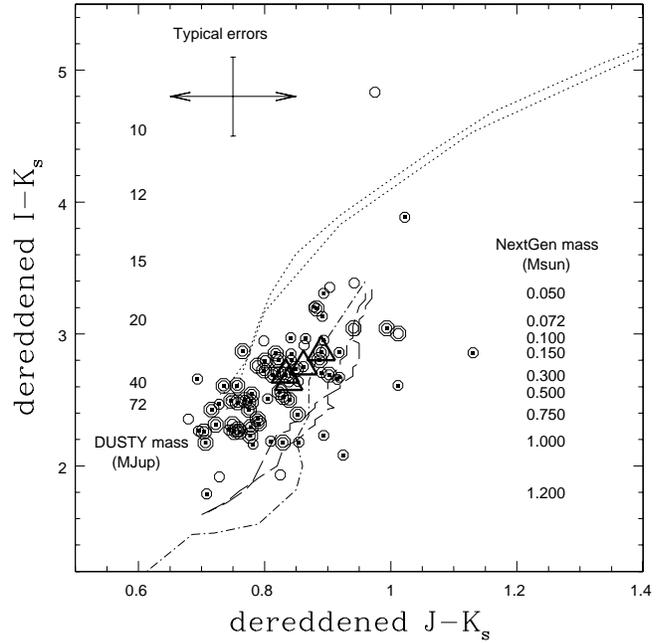}
\caption{As Fig.\,\ref{f4}, but for dereddened $I-K_s,J-K_s$. Again, mass scales are derived using the $I-K_s$ colour, with 
typical errobar shown. The error in $J-K_s$ is large, but it is seen that most candidates lie in a space quite consistent with
either NextGen or DUSTY models for the cluster (assumed distance modulus 9.4).  }
\label{f5}
\end{figure}

\begin{figure}
\centering
\includegraphics[width=9cm]{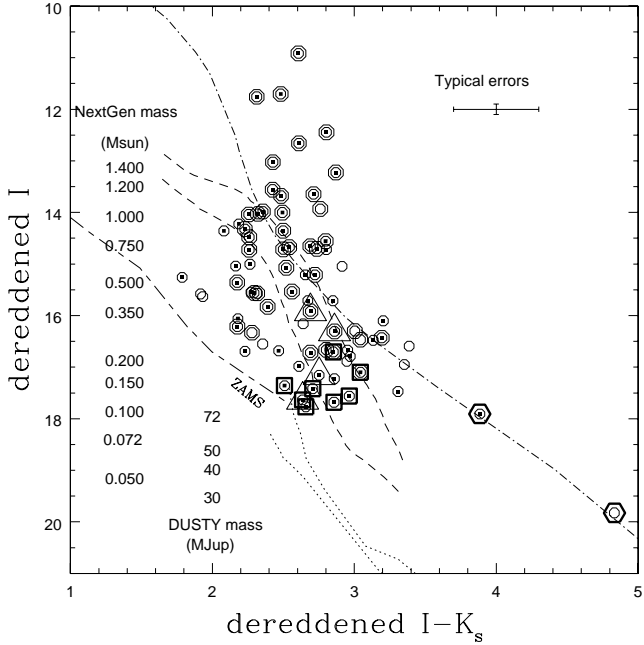}
\caption{The dereddened $I,I-K_s$ diagram. The 65 objects we have in common with the sample of Lamm et al. (2004) 
are plotted as small filled squares; objects 
with 2MASS photometry flagged AAA are overplotted as open octagons. The errorbar for $I$ is $\pm$\,0.1\,mag. which is an
overestimate for the brighter part of the sample. Note that the errorbars do not take into account uncertainties in A$_{\rm v}$; i.e. this
quantity is treated as exact. Large open triangles indicate the four previously identified NGC\,2264 members from Fig.\,\ref{f4}. 
Mass scales are shown for both NextGen and DUSTY models (dashed and dotted 
lines, respectively). The long/short dashed line is the zero-age main sequence (see text). The dash-dot line is a 5\,Gyr field model, 
chosen for distance modulus 4 (63\,pc), chosen to fit 
the 2 objects with $I-K_s$\,$>$\,3.8 which are discussed in the text. All other objects have locations with respect to the isochrones
consistent with them being cluster members with masses above 0.1\,M$_{\odot}$ (NextGen) but the existence of a few objects
very close to the substellar limit is not ruled out by the DUSTY models. The most likely such objects are overplotted by bold squares, and will
be discussed in the text. Bold hexagons highlight two further objects whose candidacy will be discussed. }
\label{f6}
\end{figure}

\begin{figure}
\includegraphics[width=9cm]{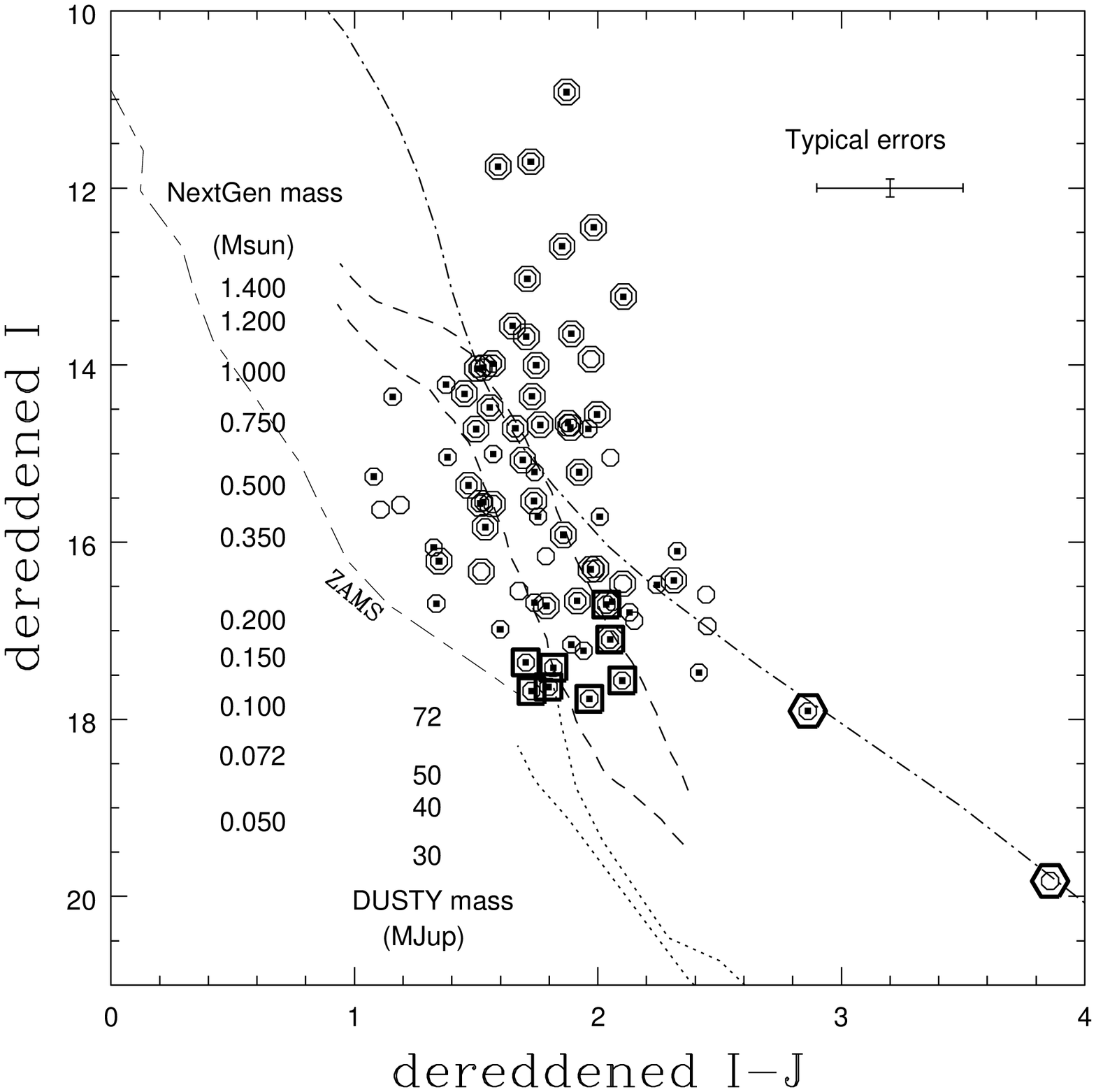}
\caption{As for Fig.\,\ref{f6}, for dereddened $I,I-J$.}
\label{f7}
\end{figure}


\section{Discussion}

In this section we will discuss and interpret the nature of our sample by consideration of their dereddened
locations in a number of colour-colour and colour-magnitude diagrams, with respect to theoretical DUSTY and
NextGen models (\cite{bar03} and references therein) and other loci derived observationally.  Initially, we have treated
our values of A$_{\rm v}$ as exact, deferring a discussion of uncertainties until Sect. 4.3. Hence in Figs.\,\ref{f4} to \,\ref{f7}
the errorbars reflect only the observed photometric uncertainties. In Fig.\,\ref{f4}
is plotted the $I-K_s$ {\it vs.} $J-H$ diagram. The $J-H$ colour is in itself not a diagnostic; all points 
lie in the range 0.4--0.8 which is comparable to the likely error in this colour from 2MASS, at least for those
objects with poorer quality photometry. However,
regarding the $I-K_s$ colour, it can be seen that a large range is covered by our points, spanning likely
solar-mass or higher pre-main sequence (PMS) stars (according to the NextGen models) all the way down to around 50\,M$_{\rm Jup}$ or
even lower, where DUSTY model predictions become appropriate. Our photometry is not accurate enough to distinguish
whether a given point lies preferentially near a DUSTY or NextGen model, at any mass where the model predictions overlap. Of course, at the 
3\,Myr age of
NGC\,2264, an object at the substellar limit would have spectral type $\sim$\,M6--M7, above the temperature
(2500--2600\,K or M9/L0) at which dust formation becomes important in the atmospheres of low-mass objects; proven association with
a DUSTY isochrone must imply a mass well below this, as the mass scales plotted in Fig.\,\ref{f4} and subsequently show.  It is clear
from Fig.\,\ref{f4} that almost all points lie at positions extremely consistent with the predictions of
NextGen isochrones. This is
borne out by a cross-correlation of the 79 objects in our sample with a much larger dataset of 10554 objects in the field
of NGC\,2264 ({\cite{lam04}), by which we recover 65 objects, plotted as small filled squares in Fig.\,\ref{f4}
and subsequently. However, it should be cautioned that by no means all these objects are
confirmed members. Further cross-correlation reveals only four confirmed 
PMS periodic variables, which are plotted as bold triangles in Fig.\,\ref{f4} and  Fig.\,\ref{f5} and also as triangles in Fig.\,\ref{f6}. 
We note we have no objects in common with the sample of Rebull et al. (2002), because their limiting magnitude is only $I$\,$\sim$\,17.9.

It is clear that the vast majority of our faint sample have colours consistent with the NextGen isochrones for NGC\,2264. 
Of course, some fraction could be foreground field stars; the dash-dot line
in Fig.\,\ref{f4} represents such a locus of observed dwarf colours given by \cite{leg92} and \cite{kir00}. 
We do not yet have data to distinguish the cluster and field objects by, for example, gravity-sensitive near-infrared spectroscopy
(\cite{luc01}).
We will return to the topic of foreground field contaminants in more detail in a later section.

Fig.\,\ref{f5} is similar to Fig.\,\ref{f4} except we have plotted the $J-K_s$ colour instead of $H-K_s$. Doing so alleviates
slightly the problem of potential large errors in the 2MASS photometry, if only because the $J-K_s$ colour has a larger
intrinsic scatter in the sample. Yet, while we are still unable to differentiate between NextGen and DUSTY model predictions
here, almost all objects are located consistently with their being NGC\,2264 members. Furthermore, a few at the low-mass
tip of the NextGen isochrones could be substellar members; clearly for the objects lying near the BD limit of the 
DUSTY isochrones, further observations are required to disentangle {\it bona-fide} substellar members from higher mass
members and field objects.

For the current study, we need to rely on the absolute magnitudes to yield a truer picture of the status of the objects
in our sample. In Fig.\,\ref{f6} we plot $I$ {\it vs.} $I-K_s$, in Fig.\,\ref{f7} $I$ {\it vs.} $I-J$. 
The zero-age main sequence (ZAMS) is plotted as a long/short dashed line for spectral types earlier than M5; we have used 
absolute $V$-magnitudes from \cite{all73} together with observed colours (\cite{ken95}; Leggett 1992; Kirkpatrick et al. 2000). 
As should be the case, 
all points lie above the ZAMS in a region where PMS stars would be located. 
In no cases are the loci of DUSTY models well populated; with the obvious reason that the 
2MASS photometry utilised here is not
sufficiently deep to probe so far below the substellar limit. Nevertheless, these plots demonstrate that our $I,z$-selected
sample, as presented here, may well contain some objects at or very close to the substellar limit, with $I$-magnitudes, after
dereddening, consistent with both low mass and cluster membership. Such objects lie close to $I$\,=\,18 and near the tip
of the DUSTY isochrones at 72\,M$_{\rm Jup}$, and may either be cluster BD or slightly more massive ($\sim$\,0.1\,M$_{\odot}$)
members according to the NextGen predictions.     

\begin{table*}
\caption[]{Observed and derived data for NGC\,2264 candidates discussed Sect 4.1 (the first eight) and Sect 4.2 (last two listed). 
Columns 7, 8 and 9 give respectively the A$_{\rm v}$ derived from near-infrared colours, and the dereddened $I$ and $J$-magnitudes. Uncertainties
in $JHK$ are taken to be $\pm$\,0.1 mag. and $\pm$\,0.3 mag. for flags A and C respectively, yielding errors in A$_{\rm v}$
$\pm$\,1\,mag. for flag A, $\pm$\,3\,mag. for flag C. These errors are carried through to the dereddened magnitudes (see text for details). }
\begin{center}
\begin{tabular}{lllllllll}
\noalign{\smallskip}
\hline
\hline
\noalign{\smallskip}
2MASS designation & $I$ & $J$ & $H$ & $K$ & A$_{\rm v}$ & der. $I$ & der. $J$ &  2MASS flag \\
\hline
2MASS J06395722+0941011 & 18.09 & 16.00 & 15.27 & 14.87 & 1.39 & 17.4\,$\pm$\,0.5 & 15.6\,$\pm$\,0.3 & AAB \\
2MASS J06400642+0944197 & 18.27 & 16.10 & 15.28 & 15.23 & 1.04 & 17.8\,$\pm$\,0.5 & 15.8\,$\pm$\,0.3 & AAC \\
2MASS J06401053+0939557 & 18.39 & 15.65 & 14.66 & 14.24 & 3.49 & 16.7\,$\pm$\,0.5 & 14.7\,$\pm$\,0.3 & AAA \\
2MASS J06410604+0949232 & 18.00 & 15.71 & 15.03 & 14.70 & 0.89 & 17.6\,$\pm$\,0.5 & 15.5\,$\pm$\,0.3 & AAB \\
2MASS J06402759+0945464 & 18.05 & 16.08 & 15.38 & 15.09 & 0.86 & 17.6\,$\pm$\,0.5 & 15.8\,$\pm$\,0.3 & AAC \\
2MASS J06405674+0938101 & 18.64 & 16.40 & 15.49 & 15.15 & 2.65 & 17.4\,$\pm$\,1.0 & 15.7\,$\pm$\,0.6 & BAC \\
2MASS J06413132+0935120 & 17.95 & 15.55 & 14.86 & 14.26 & 1.77 & 17.1\,$\pm$\,0.5 & 15.1\,$\pm$\,0.3 & AAA \\
2MASS J06401789+0941546 & 18.44 & 16.40 & 15.82 & 15.00 & 1.58 & 17.7\,$\pm$\,1.5 & 16.0\,$\pm$\,0.9 & CBB \\
\hline
2MASS J06404873+0939017 & 21.25 & 16.80 & 15.97 & 15.32 & 2.97 & 19.8\,$\pm$\,1.5 & 16.0\,$\pm$\,0.9 & CCC \\
2MASS J06411281+0945529 & 20.80 & 16.74 & 15.61 & 14.69 & 6.00 & 17.9\,$\pm$\,1.5 & 15.1\,$\pm$\,0.9 & CBB \\
\noalign{\smallskip}
\hline
\end{tabular}
\end{center}
\end{table*}

\subsection{Likelihood of NGC\,2264 membership for selected low A$_{\rm v}$ candidates}

In the following discussion, we concentrate on the eight objects listed in Table 1 and overplotted as bold squares in Figs.\,\ref{f6} and \ref{f7}.
As a group, these candidates are faint ($I$\,$\sim$\,17--18 after dereddening) and lie close to the BD limit as suggested by DUSTY
models, but with higher masses $\sim$\,0.1--0.15\,M$_{\odot}$ from the NextGen models. Clearly, all are excellent low-mass NGC\,2264 candidates.
Only one is a known variable PMS star recovered in the sample of Lamm et al. (2004), as can be seen  from Fig.\,\ref{f6}. These eight objects are
selected to have low A$_{\rm v}$, lying to the left of the sloping line in Fig.\,\ref{f3a}. Their observed $I$-magnitudes lie in the range
17.9--18.6 with $I-J$\,$\sim$\,2; one object lies on the DUSTY isochrone in Fig.\,\ref{f4} with dereddened $I-K_s$\,$\sim$\,2.8 and A$_{\rm v}$\,$\sim$\,1.6, and is
a potential $\sim$\,30\,M$_{\rm Jup}$ object by this comparison. Its dereddened $I$-magnitude and $IJK_s$ colours place it at the tip of the 
2\,Myr DUSTY isochrone in Figs.\,\ref{f6} and  \ref{f7}  
yielding a mass at the substellar limit for age 2\,Myr.
Dereddening using $IJK_s$ colours only  yields A$_{\rm v}$\,=\,3.3. This object is a good candidate 
cluster object with potentially substellar mass. Alternative explanations  are  that, by reference to a 5\,Gyr
field isochrone (not plotted), it could be a $\sim$\,0.15\,M$_{\odot}$ object at $\sim$\,220\,pc. Such is also suggested 
by it lying close to the ZAMS with spectral type M5. Presumably it is insufficiently reddened to be
a background giant, behind both NGC\,2264 itself and the more distant background dark cloud. If this object is in front of the 
cloud, its extinction can only be explained by invoking a nearby, hitherto unnoticed, source of reddening,
in the {\it foreground} of NGC\,2264. If a background giant, with $I$=17.7 after dereddening and spectral type M5, it has M$_{\rm v}$\,=\,--0.8
(Allen 1973) and $V-I$\,$\sim$\,3 (Bessell \& Brett 1988). Its distance would then be an unfeasibly large 200\,kpc. If instead it is
a reddened giant of G0 type (M$_{\rm v}$\,=\,+1.1, $V-I$\,$\sim$\,0.8) then its distance would still be $\sim$\,30\,kpc. It is highly unlikely to
be a giant at these distances, especially as it is located close to the galactic plane.  Since neither the foreground dwarf
or background giant hypothesis can explain all the observed characteristics of this object, it is most likely, in our opinion,
to be an NGC\,2264 member with mass at or close to the BD limit. We note that it has no counterpart in SIMBAD within 9\,\arcsec;
the object LBM\,2808 (\cite{lam04}) is 10\,\arcsec~ distant and much brighter ($I$\,=\,16.08). Both are clearly visible on the 
2MASS $K_s$-band images where the fainter object examined here is a clear point source. Its cc\_flag\,=\,000, so the photometry is
uncontaminated; the gal\_contam and mp\_flg are also both zero; the object is not associated with any extended 2MASS source, nor any
solar system object.    

As noted, this object has characteristics in common with seven other objects also plotted as bold squares in Figs.\,\ref{f6} and \ref{f7};
low A$_{\rm v}$ in the range 0.8--3.5 from the $J-H,H-K_s$ colours, and therefore dereddened $I$-magnitudes suggesting low  masses. 
Taken as a group, it is possible to investigate whether they might be background giants, using similar arguments to the above.
If we adopt $I$\,=\,17, after dereddening, and M$_{\rm v}$\,=\,0 which would be the case for an early K-type giant, then with 
$V-I$\,=\,1.2 (approximately K2III after Bessell \& Brett 1988), we find a distance of 44\,kpc.  The dereddened $I$-magnitudes of our sample
are generally fainter than this (Fig.\,\ref{f7}) so derived distances for the eight objects could range up to $\sim$\,70\,kpc (for
$I$\,=\,18). 
Hence, we consider background contamination 
extremely unlikely, although it cannot be absolutely
ruled out without spectroscopy.

Lastly, for these eight objects we may consider the dereddened $J$-magnitude (Table 1). For a range between 15 and 16, and adopting 
M$_{\rm v}$\,=\,0 and $V-J$\,$\sim$\,2 (again derived from the colours of an early K-type giant), we find a distance range between 25 and 
40\,kpc. While the distances are feasible, given the moderate reddening and low galactic latitude, it would seem unlikely that
these are very distant objects lying beyond both NGC\,2264 itself and the background molecular cloud. NGC\,2264 membership is much more likely,
yielding masses between the substellar limit and $\sim$\,0.2\,M$_{\odot}$ from the dereddened 
$I$ magnitude (Fig.\,\ref{f7}) using NextGen isochrones. 

\subsection{Two further cases}   

Here, we return to examine the cases of the two objects in Fig.\,\ref{f6} and \ref{f7} which  are marked by 
bold hexagons. They  appear to be preferentially
selected by their locations close to DUSTY isochrones in Fig.\,\ref{f4}, and are the only two objects with $I-K_s$\,$>$\,3.5; 
potentially they are cluster members well below the substellar limit. They are very faint, red objects 
with observed $I$\,$\sim$\,21, dereddened $I$\,$\sim$\,18 and 20 and dereddened $J$\,$\sim$\,15 and 16; their A$_{\rm v}$ are indicated by the
open squares in Fig.\,\ref{f3a}. Both are also potentially field objects, 
lying on a
5\,Gyr isochrone plotted for $\sim$\,60\,pc, chosen to select both objects from their $I$-magnitudes. 

Firstly, we will consider the faintest and reddest object, with $I-K_s$\,$\sim$\,4.8 (Fig.\,\ref{f4}) and dereddened $I$\,$\sim$\,19.8
(Fig.\,\ref{f6}). 
If a foreground dwarf, its absolute $I$ and $J$-magnitudes suggest it has a mass
between 0.075 and 0.08\,M$_{\odot}$ with respect to the 5\,Gyr  isochrone; just above the BD limit. 
From its near-infrared colours only, we derive 
A$_{\rm v}$\,=\,3, extremely unlikely for an object so relatively close.  
A$_{\rm v}$ derived from the observed $I-J,J-K_s$ colour (Fig.\,\ref{f2}) is 0.85. This lower value is also rather
unlikely for an object in close proximity, even near the galactic plane. Nevertheless, it might be  that this object is indeed a
reddened (A$_{\rm v}$\,$\sim$\,1--3) dwarf at $\sim$\,60\,pc, 
with a mass just above the brown dwarf limit, and hitherto unnoticed in the galactic plane. There is no counterpart in the SIMBAD database 
within 10\,\arcsec. 
Were it an NGC\,2264 member, its absolute
$I$-magnitude would suggest a mass of 30\,M$_{\rm Jup}$; but rather higher from the $J$-magnitude. The current data are unable to tell with certainty
whether this object is an NGC\,2264 member, or a foreground star.
However, it cannot be a background giant; it is too faint, unless it is a very massive star. 
For example, for a G0 giant with M$_{\rm v}$\,=\,+1.1
and $V-I$\,$\sim$\,0.8, then the distance is $\sim$\,80\,kpc; and an intrinsically brighter giant with earlier or later spectral type would be 
even more remote. Taking all the evidence presented here, we consider that this object is most likely to be an NGC\,2264 member with potentially
very low mass. However more accurate photometry than provided by 2MASS (this object is flagged ``C'' for all magnitudes) is needed.

We now turn to the second of these objects, with $I-K_s$\,$\sim$\,3.8 and dereddened $I$\,$\sim$\,18. Again there is no SIMBAD counterpart within 
$\sim$\,8\,\arcsec; the object referred to by Lamm et al. (2004) as LBM\,6160 (their Table 4) is within 10\,\arcsec, but is much brighter
($I$\,=\,16.14), not a confirmed cluster member according to these authors, and clearly not the same object as discussed below. For our candidate,  
both methods of calculating A$_{\rm v}$ yield a value between 5 and 6; it 
also lies along the field 5\,Gyr, 60\,pc isochrone which suggests a mass between 0.08 and 0.09\,M$_{\odot}$. It is extremely hard to see
how a nearby object could be so reddened, and similar arguments to those presented above do not provide a plausible case
for the object being a background giant; with G0 giant characteristics, its distance would be $\sim$\,30\,kpc. 
Inspection of Fig.\,\ref{f6} suggests it might be more likely to
be a cluster member with $\sim$\,0.1\,M$_{\odot}$, by reference to the NextGen isochrones, given the errorbar in $I-K_s$. Indeed, the 
2MASS $K_s$-magnitude is flagged ``B''; were the object to be in reality fainter at $K_s$, the $I-K_s$ colour would be bluer; closer to the
cluster isochrone. It would appear most likely that this object is a hitherto unrecognised, low-mass but probably not substellar, NGC\,2264 member.

As a final comment, we note that both these objects have 2MASS flags gal\_contam\,=\,0 and mp\_flg\,=\,0; thus neither are associated with objects
in the extended source catalogue, nor with known solar system objects. The brighter object of the two has a confusion flag associated with the
$J$-magnitude; not an uncommon occurence in regions of high stellar density. Presumably this might be because of the proximity 
on the sky of the Lamm et al. object LBM\,6160. Lastly, both appear as faint point sources on inspection of the 2MASS quicklook
$K_s$-band images.  

In summary, it is apparent that while the membership of these two objects in NGC\,2264 remains rather uncertain, and in spite of the fact that
they do not lie close to theoretical isochrones for this cluster, simple arguments show that it is highly unlikely that they are
either foreground dwarfs, because of their A$_{\rm v}$, or background giants, because of their low galactic latitude, faintness, and 
large derived distances. Therefore, they remain excellent candidates for NGC\,2264 membership, in which case their $I$-magnitudes
place one quite highly reddened object at $\sim$\,0.1\,M$_{\odot}$ (NextGen) and suggest the faintest is a cluster member with mass near 
30\,M$_{\rm Jup}$ (DUSTY). 

\begin{figure}
\includegraphics[width=9cm]{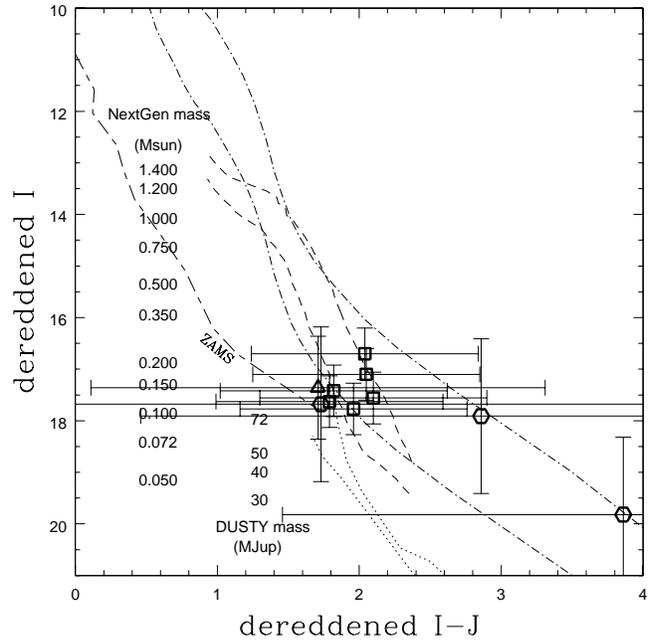}
\caption{Dereddened $I,I-J$ diagram illustrating the likely uncertainties on masses once 2MASS photometric errors are propagated through
the dereddening process. The bold squares have $J$-mag. flag A for which we adopt $\Delta$A$_{\rm v}$\,=\,$\pm$\,1\,mag., 
hence $\Delta$$I$ is $\pm$\,0.5 mag, $\Delta$$J$\,$\pm$\,0.3 mag. and $\Delta$$I-J$\,$\pm$\,0.8 mag. {\it maximum}. For the other objects (triangles;
$J$ flag B, hexagons; $J$ flag C),we adopt
$\Delta$A$_{\rm v}$\,=\,$\pm$\,2 and 3\,mag. respectively, with correspondingly larger errorbars in $I$ and $I-J$. Dash-dot lines are 5\,Gyr NextGen isochrones
for distance moduli 4 (as in Fig.\,\ref{f7}) and 6. Other isochrones are as Fig.\,\ref{f7}.}
\label{f10}
\end{figure}

\subsection{Treatment of errors in dereddened magnitudes} 

Before concluding this study, we would like to consider further these ten most probable NGC\,2264 members, in particular with a view to
investigating contamination by foreground M-dwarfs and also how uncertainties in the dereddened magnitudes translate to mass uncertainties. By
inspection of the errorbars and magnitude of the reddening vector in Fig.\,\ref{f3}, we will adopt uncertainties between $\pm$\,1 and $\pm$\,3\,mag. in A$_{\rm v}$
for the objects with 2MASS flags A and C respectively. In general, for these red (faintest at $J$) objects, poor accuracy in the 2MASS $J$-magnitude suggests that
$J$ is fainter and the $J-H$ colour is larger than given in Fig.\,\ref{f3}, and A$_{\rm v}$ correspondingly greater. 
Therefore in practice, because A$_{\rm v}$ is rather more dependent on $J-H$ than $H-K$ (Fig.\,\ref{f3}), we have used
the $J$-magnitude flag A, B or C to determine $\Delta$A$_{\rm v}$\,=\,$\pm$\,1, 2 or 3 mag., with corresponding $\Delta$$I$ and $\Delta$$J$ from the 
usual relations (\cite{rl85}), approximated
in Fig.\,\ref{f10}. Here, we repeat Fig.\,\ref{f7} plotting only the 10 objects discussed in detail above,
showing errorbars resulting from the uncertainties in dereddening. It is clear that even for a member of this set at 0.1\,M$_{\odot}$, with the largest
dereddening errors, the final derived mass is nevertheless restricted to a range from $\sim$\,0.3\,M$_{\odot}$ (NextGen) down to the substellar limit or
below, with the caveat that A$_{\rm v}$ is more likely to
be under- than over-estimated for the faintest objects and masses are therefore likely to be at the upper end of the given range. For objects
with the best $J$-band photometry, the errors on the final masses are quite small; the five objects with $J$-flag A (bold squares in Fig.\,\ref{f10}) clearly lie
within $\sim$\,0.1--0.2\,M$_{\odot}$ if cluster members.  Furthermore, the large error in the dereddened $I-J$ colour for the faintest and reddest candidate does not rule out
its being a $\sim$\,30\,M$_{\rm Jup}$ member with reference to the NextGen or DUSTY model isochrones. With regard to contamination by foreground dwarfs, the current data are
unable to rule out this possibility, and the main group of candidates have dereddened locations compatible with a 5\,Gyr isochrone at distance modulus 6 (Fig.\,\ref{f10}).
As has already been made clear, spectroscopy is required to identify young NGC\,2264 objects by their low surface gravities. It is interesting to note, though, that if
a foreground object (at 60\,pc) the faintest object in our sample is itself at the NextGen BD limit. 
In summary, given the likely uncertainties on the final masses and those inherent in the model predictions, all ten objects remain good candidates
for low-mass, NGC\,2264 members and  possibly cluster brown dwarfs, worthy of further examination.  

\begin{figure}
\includegraphics[width=9.0cm]{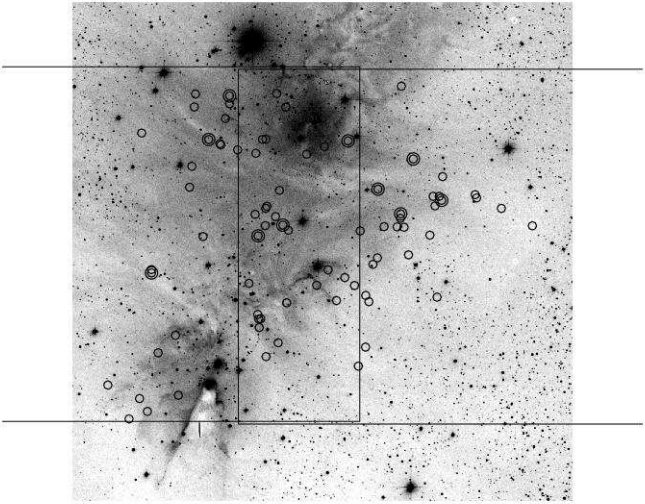}
\caption{Spatial location of all 79 candidate objects plotted
on the optical ($R$-band) DSS image which spans $\sim$\,40\,\arcmin\,$\times$\,40\,\arcmin. 
The 10 objects discussed in Sect.\,4.1 and 4.2 are overplotted as larger 20\,\arcsec~circles.
The fields of view of the two 12K fields observed are shown, and extend in RA beyond the DSS image. North is up and East to the left.} 
\label{f9}
\end{figure}

\section{Concluding remarks} 

We have presented our initial findings from a deep, wide-field CFH12K $I,z$-band survey of $\sim$\,0.6\,deg$^2$ in NGC\,2264, which 
is complete to $I$\,$\sim$\,22, several magnitudes deeper than previously published surveys. The
area surveyed is shown in Fig.\,\ref{f9}, where the locations of our sample of 79 candidate members are overplotted on the $R$-band
DSS image. By cross-correlation with 2MASS, we have been able to accurately deredden all our sample, finding A$_{\rm v}$ variable,
between $\sim$\,1 and $\sim$\,15, as expected from the clear strong and variable extinction apparent in the image of Fig.\,\ref{f9}.
Our first important conclusion is that in a study of this type, of any similar region, all near-infrared colours are required to
deduce A$_{\rm v}$ and therefore to be able to compare dereddened photometry with the predictions of model isochrones.

In doing so, we have found that all our sample have photometry consistent with their being NGC\,2264 members with masses ranging between
solar and near the substellar limit. Since all the objects have $I$\,$\ga$\,18, the most massive candidates are highly reddened, and 
we only probe close to the BD limit for low $A_{\rm v}$. We cannot at this stage rule out some contamination by background giants, 
although significant contamination by such objects is unlikely. We have examined thoroughly the cases of the reddest objects with low
A$_{\rm v}$, and find that all are potentially members with a mass very close to the substellar boundary or below; one may be a member
with significantly lower mass. Alternative explanations remain possible, but good arguments can be made to refute them. 
Infrared spectroscopy is required to finally settle the issue. 

Still, we have not yet been able to probe the substellar region with certainty, since the 2MASS survey is not quite deep enough to
provide counterparts to all 236 optical candidates; this is not surprising, since at $I$\,=\,22 and assuming zero reddening, a
substellar member might have $I-K_s$\,$\sim$\,3.5 (by reference to both observed colours for spectral type $\sim$\,M7 and NextGen isochrones).
Hence $K_s$ would be 18.5, well beyond the reach of 2MASS. Deeper near-infrared photometry is needed; thus NGC\,2264 is an excellent target
for the next generation of wide-field near-infrared surveys, including the UKIRT/UKIDSS/WFCAM survey (for which NGC\,2264 will be observed
as part of the Galactic Plane Survey), or CFHT/WIRCAM.

\begin{acknowledgements}
     TRK, EM and DJJ acknowledge support from the 5th Framework European Union Research Training Network ``The Formation and
Evolution of Young Stellar Clusters'' (RTN1-1999-00436) and TRK from the French {\it Minist\`{e}re de la Recherche}.
EM also acknowledges support from The Particle Physics and Astronomy Research Council (PPARC) of the UK. 
We acknowledge partial financial support from the ``Programme National de 
Physique Stellaire'' (PNPS) of CNRS/INSU, France. We thank also the Director of the CFHT for allocation of directors' 
time to collect part of the data presented here. This publication makes use of data products from the Two Micron All Sky
Survey, which is a joint project of the University of Massachusetts and the Infrared Processing and Analysis
Centre/California Institute of Technology, funded by the National Aeronautics and Space Administration and the
National Science Foundation. This research has made used of the SIMBAD database, operated at CDS, Strasbourg, France.
The authors thank the referee, V.J.S. B\'{e}jar, for useful comments. 
Finally, TRK thanks Jean-Louis Monin for his helpful comments on an earlier version of this work.
 
\end{acknowledgements}


\begin{thebibliography}{}

\bibitem[Allen (1973)]{all73} Allen, C.W., {\it Astrophysical Quantities}, The Athlone Press, 1973

\bibitem[Baraffe et al. 2003]{bar03} Baraffe, I., Chabrier, G., Barman, T.S., et al. 2003, A\&A, 402, 701

\bibitem[Basri 2000]{bas00} Basri, G. 2000,  ARA\&A, 38, 485

\bibitem[Bate et al. 2003]{bat03} Bate, M.R., Bonnell, I.A., \& Bromm, V. 2003, MNRAS, 339, 577

\bibitem[B\'{e}jar et al. 2001]{bej01}B\'{e}jar, V.J.S., Mart\'{\i}n, E.L., Zapatero-Osorio, M.R., et al. 2001, ApJ, 556, 830

\bibitem[Bertin \& Amouts 1996]{ber96} Bertin, E., \& Amouts, S. 1996, A\&AS, 117, 393

\bibitem[Bessell (1991)]{bes91}Bessell, M.S. 1991, AJ, 101, 622

\bibitem[Bessell \& Brett (1988)]{bb88} Bessell, M.S., \& Brett, J.M. 1988, PASP, 100, 1134

\bibitem[Brice\~{n}o et al. (2002)]{bri02} Brice\~{n}o, C., Luhman, K.L., Hartmann, L., et al. 2002, ApJ, 580, 317

\bibitem[Caballero et al. (2004)]{cab04} Caballero, J.A., B\'{e}jar, V.J.S., Rebolo, R., \& Zapatero Osorio, M.R. 2004, A\&A, 424, 857

\bibitem[Chabrier et al. 2000]{cha00} Chabrier, G., et al.,  2000, ApJ, 542, 464

\bibitem[Chabrier 2002]{cha02} Chabrier, G. 2002, ApJ, 567, 404

\bibitem[Chabrier 2003]{cha03} Chabrier, G. 2003, PASP, 115, 763

\bibitem[Cruz et al. 2003]{cru03} Cruz, K.L., Reid, I.N., Liebert, J., et al. 2003, AJ, 126, 2421

\bibitem[Greene \& Lada 1996]{gl96} Greene, T.P., \& Lada, C.J. 1996, AJ, 112, 2184




\bibitem[Kenyon \& Hartmann 1995]{ken95} Kenyon, S.M., \& Hartmann, L. 1995, ApJS, 101, 117

\bibitem[Kirkpatrick et al. (2000)]{kir00}Kirkpatrick, J.D., Reid, I.N., Liebert, J., Gizis, J.E., et al. 2000, AJ, 120, 447

\bibitem[Lamm et al. 2004]{lam04} Lamm, M.H., Bailer-Jones, C.A.L., Mundt, R., et al. 2004, A\&A, 417, 557

\bibitem[Landolt 1992]{lan92} Landolt, A.U. 1992 AJ, 104, 340

\bibitem[Leggett (1992)]{leg92} Leggett, S.K. 1992, ApJS, 82, 351

\bibitem[Leggett et al. (1998)]{leg98} Leggett, S.K., Allard, F., Hauschildt, P.H. 1998, ApJ, 509, 836 

\bibitem[Lucas et al. 2001]{luc01} Lucas, P.W., Roche, P.F., Allard, F., \& Hauschildt, P.H. 2001, MNRAS, 326, 695

\bibitem[Luhman et al. (2003)]{luh03} Luhman, K.L., Briceno, C., Stauffer, J.R., et al. 2003, ApJ, 590, 348

\bibitem[Mart\'{\i}n et al. 2001]{mar01} Mart\'{\i}n, E.L., Dougados, C., Magnier, E., et al. 2001, ApJ, 561, L195

\bibitem[Mart\'{\i}n et al. 2004]{mar04} Mart\'{\i}n, E.L., Delfosse, X., \& Guieu, S. 2004, AJ, 127, 449 

\bibitem[Moraux et al. 2003]{mor03}Moraux, E., Bouvier, J., Stauffer, J.R., \& Cuillandre, J.-C. 2003, A\&A, 400, 891

\bibitem[Moraux \& Clarke (2004)]{mor04} Moraux, E., \& Clarke, C. 2004, A\&A, accepted, (astro-ph/0410026)


\bibitem[Padoan \& Nordlund 2002]{pad02} Padoan, P.,  \& Nordlund, A. 2002, ApJ, 576, 870

\bibitem[Park et al. 2000]{par00} Park, B., Sung, H., Bessell, M., \& Kang, Y. 2000, AJ, 120, 894


\bibitem[Rebull et al. 2002]{reb02} Rebull, L.M., Makidon, R.B., Strom, S.E., et al. 2002, AJ, 123, 1528

\bibitem[Reipurth \& Clarke 2001]{rei01} Reipurth, B., \& Clarke, C.  2001, AJ, 122, 432

\bibitem[Rieke \& Lebofsky 1985]{rl85} Rieke, G.K., \& Lebofsky, M. J. 1985, ApJ, 288, 618

\bibitem[Schlegel et al. (1998)]{sch98} Schlegel, D.J., Finkbeiner, D.P., \& Davis, M. 1998, ApJ, 500, 525

\bibitem[Sterzik \& Durisen 2003]{ste03} Sterzik, M.F., \& Durisen, R,H. 2003, A\&A, 400, 1031

\bibitem[Sung et al. 1997]{sun97} Sung, H., Bessell, M., \& Lee, S.-W. 1997, AJ, 114, 2644  

 
\end{thebibliography}
\end{document}